\documentclass[twocolumn,showpacs,preprintnumbers,amsmath,amssymb,prl]{revtex4}

\usepackage{graphicx}
\usepackage{dcolumn}
\usepackage{bm}

\newcommand{\unit}[1]{\,\textrm{#1}}

\begin{document}


\title{Asymmetric magnetization reversal in the \\
exchange bias system Fe/FeF$_2$ studied by MOKE}


\author{A.~Tillmanns$^1$}
\author{S.~Oertker$^1$}\altaffiliation[Present address:]{ IBI,
Forschungszentrum J\"{u}lich, 52425 J\"{u}lich, Germany}
\author{B.~Beschoten$^1$}\email{beschoten@physik.rwth-aachen.de}
\author{G.~G{\"u}ntherodt$^1$}
\author{J. Eisenmenger$^2$}\altaffiliation[Present address:]{ Abteilung
Festk\"{o}rperphysik, Universit\"{a}t Ulm, 89069 Ulm, Germany}
\author{Ivan K. Schuller$^2$}
\affiliation{{\rm $^1\!$}II. Physikalisches Institut, RWTH Aachen,
52056 Aachen, Germany
\\
{\rm $^2\!$}Department of Physics, University of California - San
Diego, La Jolla, California 92093-0319}
\date{\today}
\begin{abstract}
The asymmetry of the magnetization reversal process in exchange
biased Fe/FeF$_2$ has been studied by magneto-optical Kerr effect.
Qualitatively different transverse magnetization loops are observed
for different directions of the cooling and the measuring field.
These loops can be simulated by a simple calculation of the total
energy density which includes the relevant magnetic anisotropies and
coherent magnetization rotation only. Asymmetric magnetization
reversal is shown to originate from the unidirectional anisotropy
and may be observed if the external measuring field is not collinear
with either the exchange bias or the easy axis of the
antiferromagnetic epitaxial FeF$_2$(110) layer.
\end{abstract}



\maketitle

A bilayer system composed of an antiferromagnetic (AFM) and a
ferromagnetic (FM) layer exhibits a shift of the hysteresis loop
along the field axis, the so-called exchange bias
(EB)~\cite{meiklejohn1957}. This shift may be observed after cooling
the system below the N\'{e}el temperature of the AFM either in an
external magnetic field or with the FM layer magnetized to
saturation. Exchange biased systems including Fe/FeF$_2$ can
additionally exhibit a pronounced asymmetry of the hysteresis loops.

This reversal asymmetry was first investigated in Fe/FeF$_2$ by
polarized neutron reflectometry (PNR)~\cite{fitzsimmons2000}
indicating different reversal mechanisms on either side of the
hysteresis loop, which have been interpreted as coherent
magnetization rotation near the left-side coercive field and as
domain wall nucleation and propagation near the right-side coercive
field. While the former mechanism was identified by a transverse
magnetization component related to strong spin-flip scattering of
the polarized neutrons, the latter was assigned by the absence of a
transverse magnetization component. This behavior raises the
question as to the origin of this unprecedented asymmetry in any
switchable hysteretic physical system.

Asymmetric magnetization reversal has also been observed in other
exchange bias systems. But the asymmetry in the loop might also be
reversed. In, e.g.\@, Co/CoO coherent rotation has been found on the
right side of the magnetization loop only~\cite{radu2003}. A fast
and powerful experimental probe of coherent magnetization rotation
is the magneto-optic Kerr effect (MOKE)
\cite{tillmanns0508635,cord2003, mewes}. Our initial MOKE
measurements on Fe/FeF$_2$ gave also evidence for coherent rotation
on only the right side of the hysteresis loop in contrast to the PNR
results. This calls for systematic studies of the reversal asymmetry
as for different directions of the in-plane cooling field relative
to the easy axis of the EB system. The angular dependence is also
aimed at investigating the role of higher order anisotropies in
asymmetric magnetization switching, especially of odd symmetry as
considered in Ref. \cite{krivorotov2002}.

Here we present a systematic MOKE study of magnetization reversal in
exchange biased Fe/FeF$_2$. Particular emphasis is given to measure
the net transverse magnetization component $M_T$, which is oriented
perpendicular to the external magnetic field. It may build up during
magnetization reversal mainly near the coercive fields and is a
direct probe of coherent magnetization
rotation~\cite{tillmanns0508635}. Furthermore, this technique allows
to determine the rotational \emph{direction} (chirality) of the
magnetization vector via the sign of $M_T$. Asymmetric magnetization
reversal can be identified by a non-zero $M_T$ at only one coercive
field during reversal. A quite unexpected behavior of $M_T$ is found
for different directions of the initial cooling field and the
subsequent measuring fields in a small angular range around the easy
axis of the antiferromagnet FeF$_2$. Within about $\pm3 ^\circ$ the
asymmetry in the transverse loops reverses, i.e.\@ $M_T$ switches
from the left to the right side of the hysteresis loop and changes
its sign. To simulate these loops, we present a simple model
describing the coherent rotation of a single magnetic moment using a
total energy density comprising fourfold, twofold, and
unidirectional anisotropies. The model reproduces all salient
features of the magnetization reversal. We demonstrate that
asymmetric magnetization reversal originates from the existence of
the unidirectional anisotropy and can only be observed if the
measuring field is non-collinear with either the easy axis direction
of the antiferromagnetic layer or the exchange bias direction set by
the cooling field. We argue that domain wall nucleation and
propagation is not relevant for the existence of asymmetric
magnetization reversal.

Polycrystalline Fe has been grown on epitaxial, twinned FeF$_2$(110)
by molecular beam epitaxy in the multilayer structure MgO(100)/
FeF$_2$(100~nm)/Fe(13~\!nm)/Al(10~\!nm), with Al as protective cap
layer. Details of the sample preparation and structural
characterization are given elsewhere~\cite{nogues1999}.

MOKE measurements have been carried out inside a magneto-optical
cryostat using a motorized sample rotator which enables sample
rotation with a precision of $\pm~0.1 ^\circ$ in an external
magnetic field aligned parallel to the film plane. For Kerr effect
we chose a reflection plane parallel to the transverse magnetization
$M_T$, i.e.\@ perpendicular to the external magnetic field. To
unambiguously detect a pure $M_T$ hysteresis loop, we chose
$s$-polarized light for the incident beam. Details of the
experimental setup are described in Ref \cite{tillmanns0508635}.
\begin{figure}[t]
\includegraphics[width=6.6cm,angle=0,clip=]{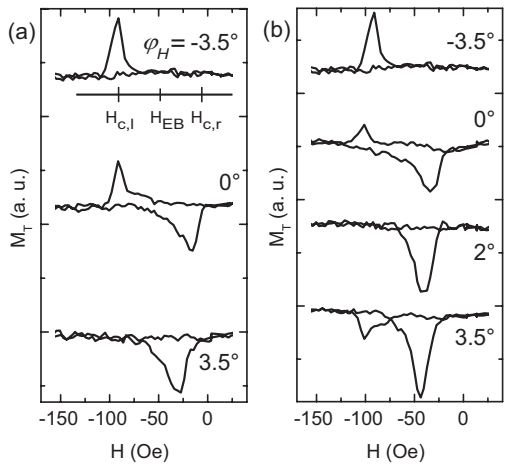}
\caption{\label{cooling} Transverse magnetization of a
polycrystalline Fe film exchange coupled to a twinned
antiferromagnetic FeF$_2$(110) layer taken by MOKE at 20~K after
field cooling in $H\! =\! 1 \unit{kOe}$ (a) at several sample angles
$\varphi_H$ at which also the measurements are performed and (b) at
-3,5$^\circ$ followed by rotations of the sample to the indicated
measurement angles.}
\end{figure}
We first investigate how the magnetization reversal depends on the
cooling field direction which we vary relative to the easy axis of
the antiferromagnetic FeF$_2$ layer by rotating the sample. Note
that for a sample angle  $\varphi_H\!=\!0^\circ$ the cooling field
is the easy axis of the AFM at 45$^\circ$ with respect to the AFM
twins along the $<$001$>$ directions. Fig.\ \ref{cooling}(a) shows a
series of transverse hysteresis loops taken at $T\!=\!20\unit{K}$
for different sample orientations (positive angles correspond to a
clockwise rotation of the sample relative to the reflection plane).
The sample is field cooled at each angle in a magnetic field of
$H\!=\!1\unit{kOe}$ through its N\'{e}el temperature
($T_N\!=\!78.2\unit{K}$) and is subsequently measured at the same
angle. At $\varphi_H\!=\!0^\circ$ the transverse loop consists of
two peaks of opposite sign close to the left and right coercive
fields, respectively, $H_{c,l}$ and $H_{c,r}$, indicating a full
360$^\circ$ coherent rotation of the magnetization vector. However,
the reversal becomes asymmetric after field cooling only slightly
away from the easy axis direction. At $\varphi_H$ \!=
\!-3.5$^\circ$, a transverse magnetization can only be detected near
$H_{c,l}$, while at +3.5$^\circ$ it appears only near $H_{c,r}$ with
opposite sign. This might explain the apparent difference between
various experimental findings in previous reports
\cite{fitzsimmons2000, radu2003}. A slight misorientation of the
cooling field direction with respect to the AFM easy axis may lead
to qualitatively different reversal asymmetries in the transverse
magnetization.

For exploring whether the observed asymmetries depend on the field
cooling procedure, we have field cooled the sample at -3.5$^\circ$
in $H\!=\!1\unit{kOe}$ and recorded transverse $M_T$ loops at
various sample angles (Fig.\ \ref{cooling}(b)). Along the easy axis
direction ($\varphi_H\!=\!0$) we again find symmetric reversal with
slightly reduced amplitude near $H_{c,l}$. The $M_T$ peak also
switches sides and sign in going to a sample orientation of
2$^\circ$. Although the angles of this switching do not exactly
match with the previous measurements (Fig.\ \ref{cooling}(a)), all
salient loop shapes are again observed. This suggests that the
existence of the rather complex magnetization asymmetry is linked to
the local anisotropies which are only weakly affected by the field
cooling procedure.
\begin{figure}[b]
\includegraphics[width=7.5cm,angle=0,clip=]{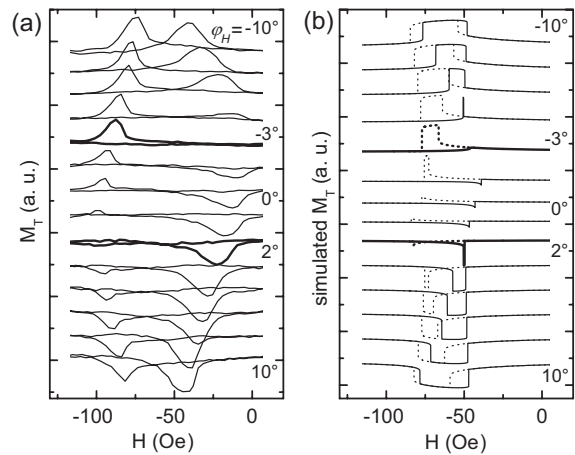}
\caption{\label{exp_sim} (a) Hysteresis loops of the transverse
magnetization of Fe/FeF$_2$ (110) at $T\! =\!20 \unit{K}$ after
field cooling along the easy axis of the AFM layer in $H \!=
\!1\unit{kOe}$. (b) Simulated $M_T$ vs. $H$ hysteresis loops for
different sample orientations using a simple model solely based on
coherent rotation (solid line: increasing field, dashed line:
decreasing field).}
\end{figure}
To further investigate how the asymmetric transverse magnetization
loops evolve we show a series of loops in a broader range of sample
angles $\varphi_H$ from -10$^\circ$ to +10$^\circ$ in Fig.\
\ref{exp_sim}(a). The data has been taken at $T \!= \!20\unit{K}$
after field cooling at 0$^\circ$ in $H \!= \!1\unit{kOe}$. At
-10$^\circ$ we clearly observe a transverse magnetization near both
$H_{c,l}$ and $H_{c,r}$ indicating symmetric reversal. Note that the
sign of $M_T$ is positive for both reversal directions in contrast
to the loop at 0$^\circ$ with opposite signs of $M_T$. A similar
reversal as at -10$^\circ$ is also seen at +10$^\circ$ with the
opposite sign of $M_T$. Most interestingly, the sign of $M_T$
reverses at different angles for both reversal directions, i.e. at
-3$^\circ$ near $H_{c,l}$ and +2$^\circ$ near $H_{c,r}$ with a
smooth change of its respective amplitudes. Note that these angles
of sign reversal mark the sample orientations of asymmetric
magnetization reversal.

We observe a transverse magnetization - which we interpret as due to
the  coherent rotation of the magnetization vector - on both sides
of the loops at almost all sample orientations. A vanishing
transverse magnetization signal on only one side of the transverse
loop is a rather exceptional case. This suggests that coherent
rotation of the magnetization may be the dominant reversal process
at all stages of reversal.

To support this scenario we use a simple model describing coherent
rotation of a single magnetic moment to (i) explain the observed
asymmetric reversal and to (ii) simulate the salient features of all
transverse magnetization loops in Fig.\ \ref{exp_sim}(a). The
reversal of the moment is induced by an external magnetic field
varied in its angular orientation. For each field step the magnetic
moment follows the minimum of a free energy density $E$ comprising
fourfold ($K_v$), twofold ($K_u$) and unidirectional ($K_e$)
anisotropies, with $E$ given by
%
\begin{alignat*}{2}
E=& -H \cos(\varphi_H)M \cos(\varphi_M) -H \sin(\varphi_H)M
\sin(\varphi_M)
\\
 &+K_v (\cos^2\!\varphi_m \cdot \sin^2\!\varphi_m)
 + K_u (
 \cos^2 [\varphi_m - (\frac{\zeta}{180}) \pi])
 \\
 &+K_e (
 \cos[\varphi_m - (\frac{\alpha}{180} ) \pi]),
\nonumber
\end{alignat*}
where $H$ describes the external magnetic field and $M$ the
saturation magnetization of the FM layer, $\varphi_H$ and
$\varphi_M$ are the angles of $H$ and $M$, respectively, which are
both measured relative to the easy axis of the AFM layer. The angle
$\zeta$ represents the respective angle of the twofold easy axis of
the FM layer and $\alpha$ the respective angle of the easy axis of
the unidirectional anisotropy $K_e$ (exchange bias). The anisotropy
constants are set to $K_{e}=\!-100000 \unit{erg/cm}^3$,
$K_{u}=\!-5000 \unit{erg/cm}^3$ and $K_{v}=\!26000 \unit{erg/cm}^3$.
These values provide the best results for the simulation of the
experimental loops in Fig.\ \ref{exp_sim}(a). The uniaxial
anisotropy constant $K_u$ of Fe is about an order of magnitude
smaller than single crystal values in literature~\cite{gradmann,
wohlfahrt}. This might be related to the large polycrystalline
fraction of the Fe film. For $M$ we use the bulk value of 1670 \!Oe
for iron~\cite{wohlfahrt}. The angle $\zeta$ is fixed to 75$^\circ$
which is approximately the easy axis direction of the ferromagnet,
while $\alpha$ is set to be identical to the cooling field direction
of 0$^\circ$ which is parallel to the easy axis direction of the AFM
layer.
\begin{figure}[t]
\includegraphics[width=6.5cm,angle=0,clip=]{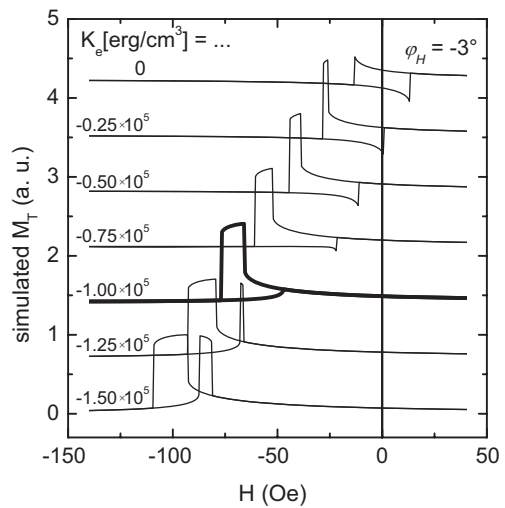}
\caption{\label{sim} Simulation of $M_T$($H$) as a function of the
unidirectional anisotropy constant $K_e$. The exchange bias
direction is assumed to be collinear with the easy axis of the AFM
layer ($\alpha\!=\!0^\circ$), while the external magnetic field is
misaligned by $\varphi_H = -3^\circ$. }
\end{figure}

We first investigate how to obtain asymmetric reversal in our
simulations. Fig.\ \ref{sim} depicts a series of simulated $M_T(H)$
loops as obtained for  $\varphi_H = -3^\circ$ and
$\alpha\!=\!0^\circ$ for various values of $K_e$. A Stoner
Wohlfarth-type magnetization reversal is observed for $K_e$ \!= \!0
(top). As expected, we observe an exchange bias shift with
increasing $K_e$. While $M_T$ remains positive near $H_{c,l}$ and
stabilizes at large values of $K_e$, it switches sign at around
$K_{e}\! =\!-1.0\times10^5 \unit{erg/cm}^3$ near $H_{c,r}$. This
clearly demonstrates that the existence of $K_e$ is solely
responsible for the asymmetry in $M_T$. We want to emphasize that we
do not obtain asymmetric reversal (not shown) if the measuring field
is collinear with both the easy axis of the AFM and the exchange
bias direction, i.e. $\varphi_H\! =\! \alpha\! =\! 0$.

We now use the above anisotropy parameters with $K_{e}\!
=\!-1.0\times10^5 \unit{erg/cm}^3$ to simulate the measured
$M_T$($H$) loops in Fig.\ \ref{exp_sim}(a). These simulations are
displayed in Fig.\ \ref{exp_sim}(a). Note that the \emph{only}
adjustable parameter is the direction of the external magnetic
field, which we change in the experiment by sample rotation. The
sign change of $M_T(H)$ on either side of the loop proceeds at the
same angles as for the measured loops thus reproducing the observed
reversal asymmetry. Differences in the shape of the simulated and
measured loops are attributed to the use of a simple macrospin model
in the simulations, which neglects any changes of the local
anisotropies that might result from the polycrystalline structure of
the Fe layer. This is evident from a more realistic Monte Carlo
simulation by Beckmann \emph{et al.}~\cite{beckmann2003}, describing
the magnetization reversal of an averaged ensemble of moments based
on the domain state model. Nevertheless, our simple macrospin model
of coherent rotation can explain all salient features of the
magnetization reversal in the experiments. Hence we conclude that
anisotropies of higher than fourth order and particularly of odd
symmetry do not play a role in describing the reversal asymmetry in
our EB system.

\begin{figure}[t]
\includegraphics[width=7.6cm,angle=0,clip=]{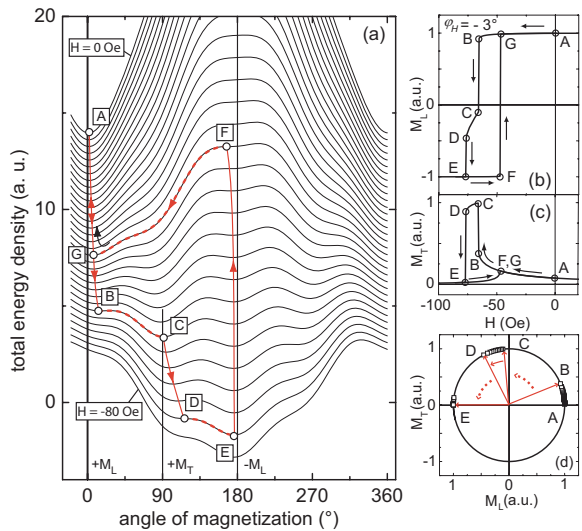}
\caption{\label{all} (a) Total energy density according to our model
(see text) as function of the external magnetic field, which varies
between 0 Oe (top) and -80 Oe (bottom). All curves are vertically
shifted for clarity. The magnetization vector follows the local
minimum of the total energy density, neglecting any thermal
activation across energy barriers. All stages of the reversal
process (A to F) are marked in (b) for the longitudinal and in (c)
for the transverse hysteresis loop. The $M_T$ vs. $M_L$ plot in (d)
illustrates the rotation of the magnetization vector along A to E
with $M_T\neq0$. }
\end{figure}

To further illustrate the origin of asymmetric reversal, we depict
the total energy density at various magnetic fields from 0\! Oe to
-80\! Oe in Fig.\ \ref{all}(a) using the parameter set of the
asymmetric loop for $\varphi_H\! =\! -3^\circ$ (Fig.\
\ref{exp_sim}(b)). Upon field variation, the magnetization vector
follows the local energy minimum. We initialize its longitudinal
orientation close to 0$^\circ$ (A) at $H = 0$ Oe. In Figs.\
\ref{all}(b) and (c), the magnetization vector is decomposed into
its $M_L$ and $M_T$ components, respectively. A polar plot of both
is included in Fig.\ \ref{all}(d). Upon reversal from A to E the
macrospin first reaches a local energy minimum near 90$^\circ$ (C)
at -65 Oe which results in a stable transverse magnetization between
points C and D (Figs.\ \ref{all}(c) and (d)), which is also observed
in the experiments (see Fig.\ \ref{exp_sim}(a)). The magnetization
is fully reversed in point E. In the opposite reversal direction the
macrospin switches between F and G with no stable intermediate state
along the transverse direction. Correspondingly, we do not observe a
stable peak on the right side of the transverse loop, although the
macrospin has to pass through the transverse direction. According to
the asymmetry of the local energy density with respect to the
direction of $-M_{L}$ at 180$^\circ$, the macrospin rotates
backwards by changing its chirality. This observation confirms the
asymmetric reversal mode simulated by Beckmann \emph{et
al.}~\cite{beckmann2003} for an angle of $\varphi_H\! =\! 60^\circ$
between the external field and the easy axis of an untwinned EB
system.

According to our simulations, the interpretation of a vanishing
component $M_T$ in experiments as a sufficient indication of
magnetization reversal by domain wall nucleation and propagation has
to be revised. Our results clearly demonstrate that conventional
domain wall nucleation and propagation is not needed to understand
the observed asymmetric magnetization reversal. However, we want to
emphasize that our simple macrospin model does not explain all
details of reversal such as the exact values of the respective
coercive fields as well as the continuous change of the net
transverse magnetization through the critical field angles at which
the sign reversal of $M_T$ is observed in the experiments. Our model
is capable of reproducing the magnetization reversal for different
cooling field directions with the same set of anisotropy constants.
The resulting phase diagram, however, is beyond the scope of the
present paper and will be reported
elsewhere~\cite{tillmannsunpublished}. A recent paper dealing with
asymmetric magnetization reversal in polycrystalline Co/IrMn relates
the asymmetry to the ratio of uniaxial FM anisotropy and EB
anisotropy as well as to the appearance of finite
coercivity~\cite{camarero}. The special case of collinear uniaxial
and unidirectional anisotropies in Ref.\cite{camarero} does not
cover the complexity of observations in our work.

In summary, we have shown that the asymmetric magnetization reversal
as probed by the transverse magnetization in Fe/FeF$_2$(110) depends
strongly on the non-zero angle between the measurement field and the
easy axis of the AFM or the EB direction. A simulation based on the
field dependent total energy density considering all relevant
anisotropies and the coherent rotation of a macrospin describes
qualitatively the experiments. The unidirectional EB anisotropy is
solely responsible for the asymmetry in the transverse magnetization
loops. The agreement between simulation and experiment endorses the
assumption that coherent rotation is sufficient to describe
asymmetric magnetization reversal in Fe/FeF$_2$.

We thank P. A. Crowell for providing the simulation code and
acknowledge useful discussions with U. Nowak. Work supported by:
DFG/SPP1133 and the US-DOE, and A. von Humboldt Foundation (I.K.S.
and J.E.). One of us (I.K.S.) thanks the RWTH faculty and
researchers for hospitality during a sabbatical stay in Aachen.

\end{document}